\documentclass[twocolumn,aps,prc,superscriptaddress,showpacs]{revtex4}
\usepackage{amsmath, bm}
\usepackage{graphicx}
\draft

\begin{document}


\title{Breaking of the number-of-constituent-quark scaling for identified-particle elliptic flow as a signal
of phase change in low-energy data taken at the BNL Relativistic
Heavy Ion Collider (RHIC)}

\author{J. Tian}
\affiliation{Shanghai Institute of Applied Physics, Chinese
Academy of Sciences, Shanghai 201800, China} \affiliation{Graduate
School of the Chinese Academy of Sciences, Beijing 100080, China}
\author{J. H. Chen}
\affiliation{Shanghai Institute of Applied Physics, Chinese
Academy of Sciences, Shanghai 201800, China}
\author{Y. G. Ma
}
\affiliation{Shanghai Institute of
Applied Physics, Chinese Academy of Sciences, Shanghai 201800,
China}
\author{X. Z. Cai}
\affiliation{Shanghai Institute of Applied Physics, Chinese
Academy of Sciences, Shanghai 201800, China}
\author{F. Jin}
\affiliation{Shanghai Institute of Applied Physics, Chinese
Academy of Sciences, Shanghai 201800, China} \affiliation{Graduate
School of the Chinese Academy of Sciences, Beijing 100080, China}
\author{G. L. Ma} \affiliation{Shanghai
Institute of Applied Physics, Chinese Academy of Sciences,
Shanghai 201800, China}
\author{S. Zhang} \affiliation{Shanghai Institute of Applied
Physics, Chinese Academy of Sciences, Shanghai 201800, China}
\affiliation{Graduate School of the Chinese Academy of Sciences,
Beijing 100080, China}
\author{C. Zhong}
\affiliation{Shanghai Institute of Applied Physics, Chinese
Academy of Sciences, Shanghai 201800, China}


\begin{abstract} We argue that measurements of identified-particle elliptic flow in a wide energy range could shed light on the
possible phase change in high-energy heavy ion collisions at the
BNL Relativistic Heavy Ion Collider (RHIC). When the hadronization
process is dominated by quark coalescence, the
number-of-constituent-quark (NCQ) scaling for the
identified-particle elliptic flow can serve as a probe for
studying the strong interacting partonic matter. In the upcoming
RHIC low-energy runs, the NCQ scaling behavior may be broken
because of the change of the effective degrees of freedom of the
hot dense matter, which corresponds to the transition from the
dominant partonic phase to the dominant hadronic phase. A
multiphase transport model is used to present the dependence of
NCQ scaling behavior on the different hadronization mechanisms.
\end{abstract}


\maketitle


The main purpose of the heavy ion program at Relativistic Heavy
Ion Collider (RHIC) in Brookhaven National Laboratory (BNL) is to
study the properties of the high density matter produced in the
heavy ion collisions, in particularly, whether it undergoes a
transition from hadronic phase to quark gluon plasma (QGP) phase.
Recently, many experimental results demonstrated that a hot and
dense partonic matter has been indeed formed in ultra-relativistic
energy heavy ion collisions at $\sqrt{s_{NN}}$ = 200 GeV~\cite{1}.
In order to study this new form of matter, probes are essential to
gain information from the earliest stage of the collisions.
Measurements of the collective motion, especially the elliptic
flow $v_{2}$ of identified particles produced in heavy ion
collisions have long been suggested as a valuable tool to study
the nature of the constituents and the equation of state of the
system in the early stage of the reaction~\cite{2}. Many
significant results in low transverse momentum range for elliptic
flow $v_{2}$ of final state particles have been obtained from RHIC
experiments~\cite{3,4,5,6,7} which consist well with the
predictions from ideal hydrodynamics~\cite{8}. More interestingly,
a number of constituent quark (NCQ) scaling behavior has been
observed in the intermediate $p_{T}$ range (1.5 GeV/c $< p_{T}<$ 5
GeV/c), which can be well reproduced by parton coalescence and
recombination model calculations~\cite{9,10,11,12}. Such scaling
indicates that the collective elliptic flow has been developed
during the partonic stage and the effective constituent quark
degree of freedom plays an important role in the hadronization
process.

Simultaneously, some experimental and theoretical developments
have suggested that important discoveries are possible at lower
collision energies where physics base may be completely different
from the pictures at higher energies. Specifically, calculations
from lattice QCD~\cite{13,14} predict a transition or fast
cross-over between the QGP state and the hadronic matter at
$T_{c}\approx $ 150$\sim$180 MeV with vanishing baryon density. In
addition, several observables in central Pb+Pb collisions from SPS
experiments show qualitative changes in the energy
dependence~\cite{15}, e.g., the full phase space ratios $\langle
K^{+} \rangle / \langle \pi^{+} \rangle$ and $(\langle \Lambda
\rangle + \langle K + \overline{K} \rangle) / \langle \pi \rangle$
showed a turnover and a decrease around $\sqrt{s_{NN}}$ $\approx$
7.6 GeV after a steep increase at lower energies.

To gain the further information about the existence of the
possible phase transition between hadronic matter and partonic
matter, the future RHIC physics program includes the  Au+Au energy
scan extending to low collision energies. Based on this, we shall
focus our studies on the NCQ-scaling of the identified particles
elliptic flow in a wide energy range. The turning on/off of the
NCQ-scaling behavior may indicate the onset of deconfinement: the
scaling will be retained in the partonic phase at high energy,
while it may be broken in lower energy when the system is
dominated by hadronic interaction.

\begin{figure} \vspace{-0.1truein}
\includegraphics[width=8.6cm]{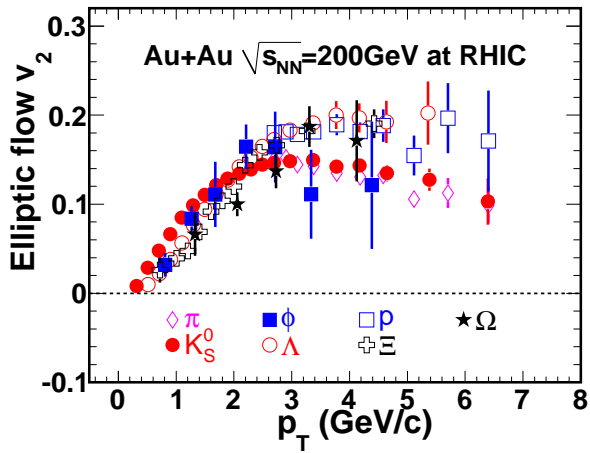}
\caption{\footnotesize (Color online) Experimental results of
elliptic flow as a function of transverse momentum for $\pi$,
$K_S^0$, $p$, $\Lambda$, $\phi$, $\Xi$, and $\Omega$ with
$|\eta|<$1.0 in 0-80$\%$
 Au+Au collisions at $\sqrt{s_{NN}}$ = 200 GeV from RHIC. The results are taken from STAR Experiments~\cite{3,4,5,6}.
}\label{fig1} \end{figure}

In the non-central collisions, the initial asymmetries in the
geometry of the system can lead to the anisotropies of the
particle momentum distributions. Since the spatial asymmetries
decrease rapidly with time, anisotropic flow can develop only in
the first few fm/c. In that way, the properties of the hot dense
matter formed during the initial stage of heavy ion collisions can
be learned by measuring the anisotropic flow. The anisotropic flow
is defined as the $n$th Fourier coefficient $v_{n}$ of the
particle distributions in emission azimuthal angle with respect to
the reaction plane~\cite{16}, which can be written as
\begin{equation} \frac{dN}{d\phi} \propto
1+2\sum v_n\cos(n\Delta\phi), \end{equation}
where $\Delta \phi$
denotes the angle between the transverse momentum of the particle
and the reaction plane. The second Fourier coefficient $v_{2}$
represents the elliptic flow which characterizes the eccentricity of
the particle distributions in momentum space. At a given rapidity
window the second coefficient is
\begin{equation}
v_{2}=\langle\cos(2\Delta\phi)\rangle=\langle\frac{p_x^2-p_y^2}{p_x^2+p_y^2}\rangle.
\end{equation}
The measured elliptic flow $v_{2}$ as a function of transverse
momentum $p_{T}$ from the minimum bias Au+Au collisions at
$\sqrt{s_{NN}}$ = 200 GeV for $\pi$, $K_{S}^{0}$, $p$, $\Lambda$,
$\Xi$, $\phi$, $\Omega$ are presented in FIG.~\ref{fig1}. The
results show that elliptic flow $v_{2}$ of all the particles
increase with the transverse momentum while the elliptic flow
$v_{2}$ becomes saturated in the intermediate transverse momentum
region. Moreover, the baryons saturate at $p_{T}$ $ \geq$ 3 GeV/c
with $v_{2}$ $\sim$ 0.2, while the mesons saturation starts
earlier at lower values of $v_{2}$. In addition, from
FIG.~\ref{fig1} we can find that although the multi-strange
baryons $\Xi$ and $\Omega$ tend to undergo less re-scatterings in
the hadronic stage, their $v_{2}$ values are as high as others
hadrons at given $p_{T}$ range, this means that the collectivity
should be developed at the partonic stage.

\begin{figure} \vspace{-0.1truein}
\includegraphics[width=8.6cm]{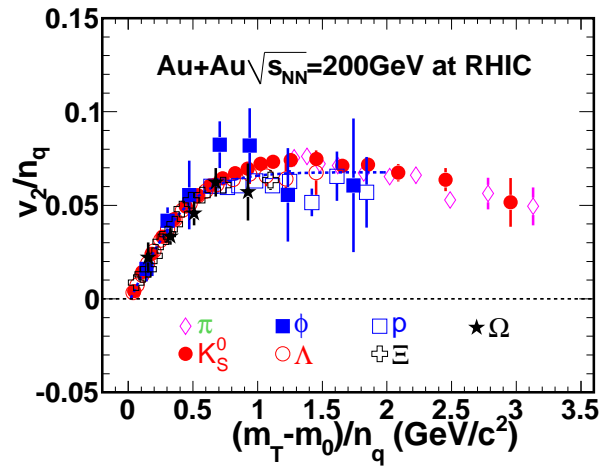}
\caption{\footnotesize (Color online) Number of constituent quark
$(n_{q})$ scaled $v_{2}$/$n_{q}$ versus scaled $(m_{T}-m)$/$n_{q}$.
Dashed blue line is the result of fit with function (3).
}\label{fig2}
\end{figure}

More intriguing phenomenon is that after scaling both values of
$v_{2}$ and $p_{T}$ by the number of the constituent quarks of the
corresponding hadron, all particles can fall onto one single
curve. However, the pion elliptic flow data are somewhat not
following the scaling, which may be caused by the large
contribution of pion yield from resonance decays~\cite{17,18}.
FIG.~\ref{fig2} shows elliptic flow $v_{2}$ as a function of
transverse kinetic energy $(m_{T}-m)$ for the identified particles
($m_{T}=\sqrt{m^{2}+p_{T}^{2}}$ is the transverse mass while $m$
is the rest mass of the particle), where $v_{2}$ and $(m_{T}-m)$
have been scaled by the number of constituent quarks ($n_{q}$).
The measured elliptic flow values can be fitted with the
equation~\cite{18} given as
\begin{equation}
f_{v_{2}}=\frac{a}{1+\exp(-(x-b)/c)}-d,
\end{equation}
where $a$, $b$, $c$ and $d$ are the parameters fixed from the fit.
This universal curve represents the momentum space anisotropy of
constituent quarks prior to hadron formation. However, this simple
scaling neglects possible higher harmonics and possible differences
between light and heavier quark flow.

\begin{figure*} \vspace{-0.1truein}
\includegraphics[width=17.2cm]{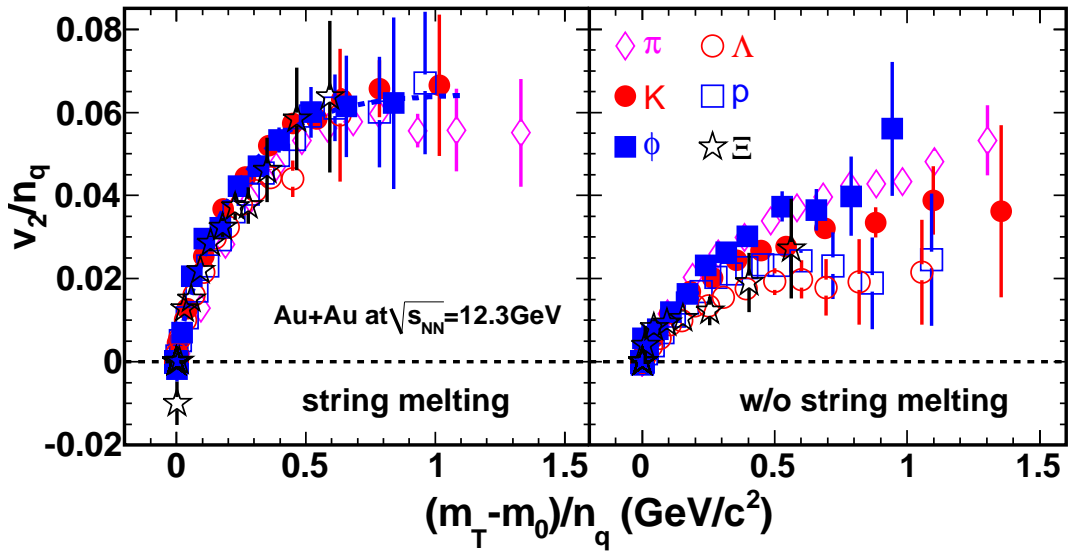}
\caption{\footnotesize (Color online) (Left panel) Number of
constituent quark ($n_{q}$) scaled $v_{2}$/$n_{q}$ versus scaled
($m_{T}-m$)/$n_{q}$ at Au+Au $\sqrt{s_{NN}}$= 12.3 GeV from the
AMPT model with string melting. Dashed blue line is the result of
fit with the function (3). (Right panel) Number of constituent
quark ($n_{q}$) scaled $v_{2}$/$n_{q}$ versus scaled
($m_{T}-m$)/$n_{q}$ at Au+Au $\sqrt{s_{NN}}$= 12.3 GeV from the
AMPT model without string melting. }\label{fig3}
\end{figure*}

Some coalescence or recombination models~\cite{9,10,11,12} can
successfully describe the hadron production in the intermediate
$p_{T}$ region, where the NCQ-scaling behavior has been observed.
According to these models, the hadronization is dominated by
coalescence of quarks, and the essential degrees of freedom seem to
be effective constituent quarks that have developed a collective
elliptic flow during the partonic evolution. There are some common
features in the intermediate $p_{T}$ region below 5 GeV/c~\cite{19}:
(i) The production probability for a baryon or meson is proportional
to the product of local parton densities for the constituent quarks;
(ii) Baryons with transverse momentum $p_{T}$ are mainly formed from
quarks with transverse momentum $\sim$ $p_{T}$/3, whereas mesons at
$p_{T}$ are mainly produced from partons with transverse momentum
$\sim$ $p_{T}$/2. Then the meson elliptic flow $v_{2,M}$ and baryon
elliptic flow $v_{2,B}$ can be given by those of partons
via~\cite{20}

\begin{equation}
 v_{2,M}(p_{T})\approx 2v_{2,q}(\frac{p_{T}}{2}),
\end{equation}
\begin{equation}
 v_{2,B}(p_{T})\approx 3v_{2,q}(\frac{p_{T}}{3}).
\end{equation}
Consequently, the collectivity of the constituent quarks become the
collectivity of the hadrons via quark coalescence during the
hadronization. Therefore, by scaling the observed $v_{2}$ signal and
the transverse momentum with the number of constituent quarks
$n_{q}$, one obtains the underlying quark flow.

Many experimental results~\cite{15} have shown anomalous
dependence on the collision energy in the SPS energy range,
especially the sharp changes around $\sqrt{s_{NN}} \approx $ 7.6
GeV. All those phenomena suggest that a common underlying physics
process is responsible for these changes. Ref.~\cite{21,22} argue
that this anomaly is probably caused by the modification of the
equation of state in the transition region between confined and
deconfined matter. The low energy runs at RHIC offer us
opportunities to study the possible phase transition, thus the
NCQ-scaling behavior for the identified particles may be
considered as a new signal of the onset of deconfinement located
in the low energy domain. We suppose that at lower collision
energies, there would be no phase transition from hadronic matter
to partonic matter. In this case, the hadrons are not produced
from the coalescence of deconfined partons, and the collective
flow does not stem from partonic stage, therefore the NCQ-scaling
behavior for identified particles will be broken.


In the following, we will use a multi-phase transport (AMPT)
model~\cite{23} to investigate the NCQ-scaling behavior of the
identified particles under different hadronization mechanisms in
Au+Au collisions. The AMPT model has an extensively agreement with
the RHIC data and AGS data~\cite{23,24,25,26,27,27b}. For example,
by using parton-scattering cross sections of 6-10 mb, it was able
to reproduce both the centrality and transverse momentum (below 2
GeV/c) dependence of the elliptic flow and pion interferometry
measured in Au+Au collisions at $\sqrt{s_{NN}}$ = 130 GeV. It can
explain the measured $p_T$ dependence of both $v_{2}$ and $ v_{4}$
of midrapidity charged hadrons and $v_{2}$ for $\phi$ meson as
well as di-hadron correlation in the same collisions at
$\sqrt{s_{NN}}$ = 200 GeV. It was also demonstrated that there
exists the NCQ-scaling of $v_2$ not only for $\pi$, $k$ and $p$
but also for multi-strange hyperons in the AMPT model with the
string melting scenario \cite{26,27b}. The AMPT model is a hybrid
model which contains many processes of Monte Carlo simulation.
There are four main components in the model: initial conditions,
partonic interactions, conversion from partonic matter into
hadronic matter, and hadronic interactions in collision evolution.
First the hybrid model uses the minijet partons from the hard
processes and the strings from the soft processes in the HIJING
model as initial phase~\cite{28}, then the dynamical evolution of
partons are modeled by the ZPC~\cite{29} parton cascade model,
which calculates two-body parton scatterings using cross sections
from pQCD with screening masses. The transition from the partonic
matter to the hadronic matter is based on the Lund string
fragmentation model~\cite{30}. The final-state hadronic
scatterings are modeled by the ART model~\cite{31}. In the default
AMPT model, minijets coexist with the remaining part of their
parent nucleons, and together they form new excited strings, then
the resulting strings fragment into hadrons according to the Lund
string fragmentation. In the AMPT model with the string melting
scenario, these strings are converted to soft partons, and their
hadronization is based on a naive quark coalescence model.

We calculated the elliptic flow $v_{2}$ with different versions of
AMPT model in Au+Au collisions at $\sqrt{s_{NN}}$ = 12.3 GeV,
which assumed to be running at RHIC in the future. Based on the
momentum of identified particles, we extract the elliptic flow
$v_{2}$ with formula (2). According to the discussion above, when
there is a phase transition from hadron matter to quark gluon
plasma, the essential degrees of freedom at the hadronization seem
to be effective constituent quarks, then we use the AMPT model
with string melting to calculate the elliptic flow. Left panel of
FIG.~\ref{fig3} shows the elliptic flow $v_{2}$ for identified
particles calculated with string melting scenario, it seems that
the elliptic flow $v_{2}$ of the identified particles have a nice
NCQ-scaling behavior due to the coalescence of partons in the
hadronization. On the other hand, if the degree of freedom of the
system is dominated by the hadronic interaction, we will use the
default AMPT model to calculate the elliptic flow, right panel of
FIG.~\ref{fig3} shows the elliptic flow $v_{2}$ of identified
particles without string melting, the NCQ-scaling of elliptic flow
$v_{2}$ is broken since hadrons are directly from the
fragmentation of excited strings. In comparison with the string
melting case, the $v_2$ values of identified particles show
significant decreasing in the case without the string melting,
which reflects that partonic stage makes an important contribution
on the development of sizable elliptic flow in the early stage.


In summary, we proposed to use the NCQ-scaling for identified
particles elliptic flow as a unique observable to probe the
effective degree of freedom of the system created in heavy ion
collisions at RHIC. We argued that the presence or absence of the
NCQ-scaling behavior might indicate the turn on or turn off of the
partonic degree of freedom of the system. The AMPT simulation has
further confirmed our argument: the NCQ-scaling is observed in the
string melting scenario where full partonic evolution has been
included while it is broken in the default version with dominant
hadronic interaction only. The upcoming RHIC low energy
measurement will help to disentangle the different physics
scenario as discussed.

This work was supported in part the National Natural Science
Foundation of China under Grant No. 10610285, 10775167 and
10705044, and the Knowledge Innovation Project of the Chinese
Academy of Sciences under Grant No. KJCX2-YW-A14 and and
KJCX3-SYW-N2.

 \bigskip
\bigskip \par

\bigskip
\end{document}